\documentclass[traditabstract]{aa} 

\usepackage{graphicx}
\usepackage{epsfig}
\usepackage{natbib}

\usepackage{txfonts}
\begin{document}
   \title{Resolved [CII] emission in a lensed quasar at z=4.4\thanks{Based on observations made with the IRAM Plateau de Bure Interferometer.
}
}

   \author{S. Gallerani\inst{1,2}
   \and
   R. Neri\inst{2}
   \and
   R. Maiolino\inst{1,3}
   \and
   S. Mart\'in\inst{4}
   \and
   C. De Breuck\inst{4}
   \and
   F. Walter\inst{5} 
   \and
   P. Caselli\inst{6}
   \and
   M. Krips\inst{2}
   \and
   M. Meneghetti\inst{7,8}
   \and
   T. Nagao\inst{9}
   \and
   J. Wagg\inst{4}
   \and
   M. Walmsley\inst{10,11}
          }

\institute{
	   INAF-Osservatorio Astronomico di Roma, via di Frascati 33, 00040
           	   Monte Porzio Catone, Italy
	\and	
           Institut de RadioAstronomie Millim\'etrique, 300 rue de la Piscine, Domaine Universitaire, 38406, Saint Martin d'H\'eres, France
        \and
	Cavendish Laboratory, University of Cambridge, 19 J. J. Thomson Ave., Cambridge CB3 0HE, UK
        \and
       European Southern Observatory, Alonso de C\'ordova 3107, Vitacura, Casilla 19001, Santiago 19, Chile  
        \and
        Max-Planck-Institut f\"ur Astronomie, K\"onigstuhl 17, 69117 Heidelberg, Germany 
	\and
           School of Physics and Astronomy, University of Leeds, Leeds LS2 9JT, UK
	\and
	   INAF-Osservatorio Astronomico di Bologna, via Ranzani 1, 40127
	   Bologna, Italy
	\and
	   INFN, Sezione di Bologna, viale Berti Pichat 6/2, 40127 Bologna, Italy
 \and
	   The Hakubi Project, Kyoto University, Yoshida-Ushinomiya-cho, Sakyo-ku, Kyoto 606-8302, Japan
	\and
	   Osservatorio Astrofisico di Arcetri,
	   Largo E. Fermi 5, 50125 Firenze, Italy
\and
Dublin Institute of Advanced Studies, 31 Fitwilliam Place, Dublin 2, Ireland
             }

   \date{Received ; accepted }

 
  \abstract
   {We present one of the first resolved maps of the [CII]~158~$\mu$m line,
a powerful tracer of the star forming inter-stellar medium, at high redshift.
We use the new IRAM PdBI receivers at 350 GHz to map this line in
BRI 0952-0115, the host galaxy of a lensed quasar at z=4.4 previously found
to be very bright in [CII] emission. The [CII] emission is clearly
resolved and our data allow us to resolve two [CII]
lensed images associated with the optical quasar images. We find that the star formation, as traced by [CII], is distributed over a region of about 1~kpc in size near the quasar nucleus, and we infer a star
formation surface density $\gtrsim$~150 $M_{\odot} ~\rm yr^{-1}~\rm kpc^{-2}$, similar to that observed in local
ULIRGs.
We also reveal another [CII] component, extended over $\sim$12~kpc, and located at about 10~kpc from the quasar.
We suggest that this component is a companion disk galaxy, in the
process of merging with the quasar host, whose rotation field
is distorted by the interaction with the quasar host,
and where star formation, although
intense, is more diffuse.
These observations suggest that galaxy merging at high-z can enhance
star formation at the same time in the form of
more compact regions, in the vicinity of the accreting black hole, and in more extended star forming galaxies.}
   {}


   \maketitle
%

\section{Introduction}

The $^2P_{3/2} \rightarrow {^2P_{1/2}}$ fine structure line of [CII] at 157.74 $\mu$m is primarily emitted by gas exposed to ultraviolet radiation in photo dissociation regions (PDRs) and is one of the major coolants of the star forming inter-stellar medium (ISM). It is the strongest emission line in most
galaxies, accounting for as much as $\sim 0.1-1$\% of their far-infrared (FIR) luminosity \citep[][]{crawford85,stacey91,wright91,stacey10,gracia-carpio11}. Given its strength, the [CII] line is in principle the most
suitable tracer to investigate the star forming ISM in galaxies and to
identify galaxies through the cosmic epochs with (sub-)mm facilities \citep[e.g.][]{maiolino08}.

While in the local Universe the [CII] line is observable only from space or airborne telescopes, at high redshift ($z\gtrsim 1$) it is shifted into the mm/submm windows, and is therefore detectable with ground-based observatories. The first detection of the [CII] line at high redshift was obtained with the IRAM 30m in
SDSS J1148+5251, one of the most distant quasars known (z=6.4),
for which the [CII] line is shifted to 1.2~mm \citep{maiolino05}. Since then, an
increasing number of [CII] detections at high redshift has been reported,
currently for a total of about 20 galaxies, with typical [CII] luminosities $\rm L_{[CII]}\sim 4.5\times 10^9-1.3\times 10^{11} \rm L_{\odot}$ (Iono et al., 2006,
Maiolino et al., 2009, Hailey-Dunsheath et al., 2010, Ivison et al., 2010, Wagg et al., 2010, Stacey et al., 2010, Cox et al., De Breuck et al., 2011, Valtchanov et al., 2011). 

By comparing the observed [CII]/FIR and [CII]/CO luminosity ratios with PDR models, it has been possible to constrain the physical conditions of the ISM in star forming regions of high redshift galaxies and, more
specifically, the intensity of the stellar radiation field, the gas
density and the chemical enrichment (Maiolino et al., 2005, Hailey-Dunsheath et al., 2010, Ivison et al., 2010, De Breuck et al., 2011, Stacey et al., 2010). 
In the local Universe, the [CII]/FIR luminosity ratio drops by an order of magnitude for sources with $\rm L_{FIR} > 10^{11}-10^{11.5} \, L_\odot$ (luminous infrared galaxies, LIRGs) \citep[][]{malhotra01,luhman98,luhman03,negishi01}. This trend has been ascribed to various possible effects, such as reduced heating efficiency of the gas in the high UV radiation fields, or the presence of a non-PDR contribution to the FIR emission, such as might arise from dust-bounded HII regions (Kaufman et al., 1999; Luhman et al., 2003; Graci\'a-Carpio et al., 2011).
 \begin{figure*}
   \centering
   \includegraphics[width=17truecm]{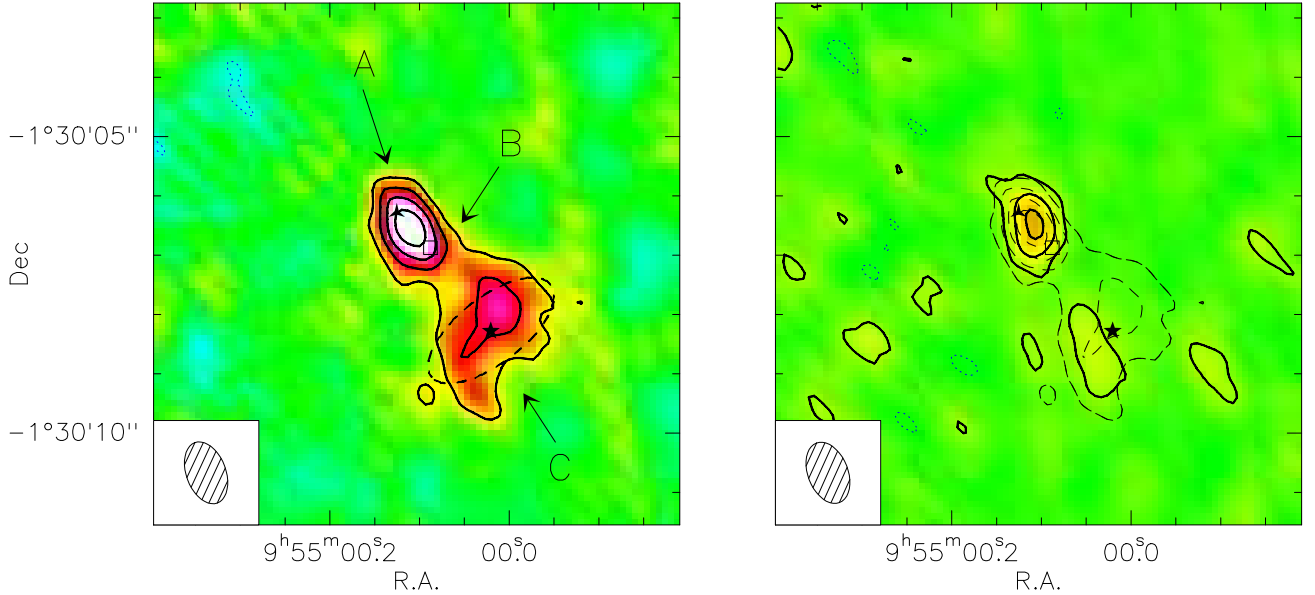}
      \caption{
Map of BRI 0952-0115 obtained with the PdBI. The synthesized beam of $1.08''\times 0.66''$ is shown
in the bottom-left insets. Left panel: map of the [CII] line emission region integrated
over a velocity range of 300 km s$^{-1}$,  i.e. $\rm -210 < v < 90~[km~s^{-1}]$. Contour levels are shown in steps of 2.5$\sigma$, where 1$\sigma=0.5$
Jy~km~s$^{-1}$~beam$^{-1}$. The filled three-points star, empty square, and filled five-points star represent the positions of
the components A, B and C, respectively (Table \ref{tabcomp}). The dashed ellipse denotes the result of the fit
to the extended component C. 
Right panel: map of the continuum emission obtained from the line-free channels of the 3.6~GHz wide spectrum (i.e. $\rm v<-210~km~s^{-1}$ and $\rm v>90~km~s^{-1}$).
Contour levels are shown in steps of 2.5$\sigma$, where 1$\sigma=0.5$ mJy~beam$^{-1}$. Dashed lines show the positive contours of the [CII] emission as in the left panel. In both panels, negative contour levels at 2.5$\sigma$ are
shown with dotted lines.} 
         \label{map}
   \end{figure*}
Mapping the [CII] emission at high redshift can provide key
information on the spatial distribution of star formation and on the
dynamics 
of primeval systems.  However, up till the present, it has only been possible to derive estimates of the size of the [CII] emitting area towards the host galaxy of the quasar SDSS J1148+5251 at z=6.4 (Walter et al., 2009).
In this object, observations with the IRAM interferometer (PdBI) have marginally resolved the [CII]
emission and the underlying continuum over an area of about one kpc in size, indicating
a star formation surface density of about 1000 $\rm
M_{\odot}~yr^{-1}~kpc^{-2}$, similar to that observed in the center
of some local ULIRGs (e.g.\ Arp220), but about two orders of magnitude
larger in area (Walter et al., 2009). Most likely, such a dense and powerful starburst is
tracing the formation of a massive spheroid in the early Universe. 
\\
In this work, we exploit the new 277-371 GHz band of the PdBI to map the [CII] emission in the host galaxy of the quasar BRI 0952-0115. This object was discovered 
by McMahon et al. (1992), and initially identified as a $z=4.5$ quasar pair separated by almost 1$''$. 
After more detailed investigations, the system has been recognized as a lensed galaxy, magnified by a 
foreground elliptical at $z=0.632$ (Leh\'ar et al., 2000; Eigenbrod et al., 2007). The detection of BRI 0952-0115 in CO(5--4) by \cite{gui99} provided an accurate determination of its redshift (z=4.4337) and an estimate of the molecular hydrogen mass ($\rm M_{H_2}\sim 2-3\times 10^9~\rm M_{\odot}$). Recently, strong [CII] emission has been discovered towards 
this object using APEX (Maiolino et al., 2009). With a peak intensity of 150$\pm$30 mJy this is one of the brightest [CII] lines detected at $z>4$ so far.
Millimeter and sub-millimeter observations indicate an apparent far-IR luminosity $\rm
L_{FIR}\sim 10^{13} \, L_\odot$ for BRI 0952-0115 \citep[]{priddey01}. However, the [CII]/FIR luminosity ratio ($\rm log
(L_{[CII]}/L_{FIR}) \approx -2.9$, Maiolino et al. 2009) is higher than the value observed in local LIRGs and ULIRGs with similar infrared luminosities.
Additional [CII] observations at high-z have confirmed the existence of a large population of distant galaxies that,
although having $\rm L_{FIR}>10^{12}~\rm L_{\odot}$, are characterized by high [CII]/FIR ratios (Stacey et al., 2010; Ivison et al.,
2010; De Breuck et al., 2011, Valtchanov et al., 2011). These high [CII]/FIR - high $\rm L_{FIR}$ - high
redshift galaxies are most likely extended, scaled up versions of local ``normal'' starbursts. However, it is possible that lower ISM metallicities also contribute to the [CII] enhancement in high redshift systems as compared to
local systems with high FIR luminosity (De Breuck et al., 2011). 

\section{Observations}\label{secobs}
\begin{figure*}
   \centering
   \includegraphics[width=12truecm]{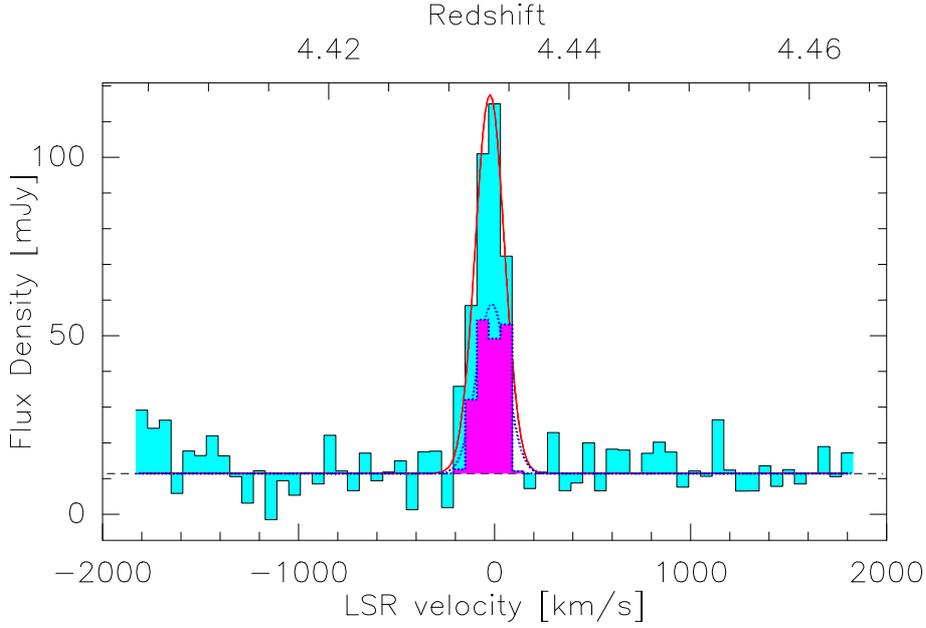}
      \caption{[CII] spectrum of BRI 0952-0115
obtained with the PdBI by integrating over all components (cyan shaded region) and over the A+B components (magenta shaded region). The spectrum is rebinned to a resolution of 60~km~s$^{-1}$ (70 MHz).}
         \label{spectrum}
   \end{figure*}

BRI 0952-0115 [(J2000) RA$=$09:55:00.1, DEC$=$--01:30:07.1] was observed with the IRAM six-element interferometer (PdBI) in 3 observing runs (Jan 15, Jan 16, Jan 17, 2011) in the extended B configuration, 
and in 2 observing runs (Mar 11, Mar 20, 2011) in the compact C configuration.  We tuned the receivers to 349.77 GHz, which is the frequency of the [CII]157.741~$\mu$m
line at redshift z=4.4337 \citep{gui99}.\\
The beam size resulting from the B+C configurations using natural weighting is $1.08''\times 0.66''$, with a position angle of 20 degree.
The $1\sigma$ sensitivity achieved by the observations is $0.5$~Jy~km~s$^{-1}$~beam$^{-1}$ in the 350 MHz (i.e. 300 km~s$^{-1}$) channel associated with the velocity-integrated [CII] emission map (see Fig. \ref{map} and Fig. \ref{spectrum}), and $0.5$~mJy~beam$^{-1}$ in the 3.2 GHz (i.e. 3300 km~s$^{-1}$) line-free channels used to determine the continuum. 
We also obtain a higher resolution map, characterized by a
synthesized circular beam of 0.5$''$ and 1$\sigma$ sensitivity of $0.8$~Jy~km~s$^{-1}$~beam$^{-1}$, by weighting the visibilities with an inverted gaussian taper.\\ 
The spectral setup provides a velocity coverage of $\sim 3600~\rm{km}~\rm{s}^{-1}$, i.e. $\sim 3.6$ GHz. A total time of 7.1 hours were spent on-source. The weather conditions were acceptable with precipitable water vapor $\rm 1.5< pwv <6.0 ~mm$, which correspond to a typical atmospheric transmission $\sim 0.4-0.8$. We adopt as flux calibrators the following sources: 3C84, 3C273, MWC349. The absolute flux calibration uncertainty is 20\%. The data were analyzed by using
CLIC and MAPPING (within the GILDAS-IRAM package\footnote{http://www.iram.fr/IRAMFR/GILDAS}).
 \begin{table*}
\caption{Properties of the [CII] line observed toward BRI 0952-0115 with PdBI (this work) compared with the
results obtained with APEX (Maiolino et al., 2009) and with the CO(5--4) line observed with PdBI \citep{gui99}. Uncertainties on the velocity integrated fluxes represent statistical errors; for the [CII] lines we report in parenthesis the total errors, which take into account also the absolute calibration uncertainties.}
\label{obs}     
\centering      
\begin{tabular}{l c c c c c c } 
\hline\hline                
Line  & Instrument&$\rm \nu _{rest}$ & $\rm \nu _{obs}$ & $z_{\rm line}$ & 
   $\rm{FWHM}$ & I  \\
      & &[GHz]&[GHz]   & & [$\rm km~s^{-1}$] &
       [$\rm Jy~km~s^{-1}$] \\
\hline 
[CII] ($\rm ^2P_{3/2}-^2P_{1/2}$)&PdBI&1900.54&349.776&4.4336$\pm$0.0001&175$\pm$11& 20.1 $\pm$ 1.3 (4.2)\\
  
[CII] ($\rm ^2P_{3/2}-^2P_{1/2}$)&APEX&1900.54&349.776&4.4336$\pm$0.0003&193$\pm$32 & 33.6$\pm$4.9 (7.0)\\

CO (5--4) & PdBI&576.2679 & 106.055  & 4.4337$\pm$0.0006 &  230$\pm$30  & 0.91$\pm$0.11 \\

\hline
\end{tabular}\\
\end{table*} 
\section{Results}
The PdBI map reveals a surprisingly complex structure (Fig. \ref{map}, left panel). 
This map has been obtained by integrating the line over a velocity range of 300 km s$^{-1}$, i.e. covering
the bulk of the line emission at $\rm -210 < v < 90~[km~s^{-1}]$. We note that the line is slightly skewed towards negative velocities. The [CII] map indicates that BRI 0952-0115 is constituted by a compact emitting region (labelled A+B in the figure) and a second more
extended component (labelled C), located about 2$''$ South-West of the A+B region.

The continuum map (Fig. \ref{map}, right panel) has been obtained by integrating the line-free channels (i.e. $\rm v<-210~km~s^{-1}$ and $\rm v>90~km~s^{-1}$). The comparison of the two maps shows that the continuum emission and the [CII] line have the same peak positions, but display a different morphology. 

Fig.~\ref{spectrum} shows the resulting spectrum, rebinned to a spectral resolution of 60~km\,s$^{-1}$ (70 MHz), and obtained by integrating over all components (cyan shaded region). The [CII] line is shown on top of a continuum of $F_{cont}=11.4\pm 1.4$~mJy. 
The line was fitted with a single gaussian, centered at $z=4.4336\pm 0.0001$ ($\nu_0=349.8$ GHz), having a
FWHM=$175\pm 11$~km~s$^{-1}$, and a peak intensity $F^{peak}_{[CII]}=106.9\pm 6.1$~mJy. 
The [CII] line center is consistent within 1$\sigma$ with that of the CO(5--4) line detected by \cite{gui99} and with the previous APEX observations by Maiolino et al. (2009).
The [CII] width is also consistent, within 1$\sigma$, with the [CII] results by Maiolino et al. (2009),
while the CO(5--4) line is 25\% wider than observed by us, but still consistent within 2$\sigma$. The PdBI flux is 40\% lower than the APEX single dish flux, although the two are consistent within 1.7$\sigma$ once the absolute calibration errors are taken into account. However, if the discrepancy is confirmed with higher accuracy, this may suggest that we are missing part of the line flux distributed on scales larger than 1$''$. The latter explanation is also supported by the fact that the PdBI continuum is $\sim$20\% smaller than the one obtained through SCUBA observations (McMahon et al. 1999), although consistent at $\sim 1\sigma$. All the results are summarized in Table \ref{obs}.\\
In Fig. \ref{spectrum}, we also plot the bulk of the line emission obtained by integrating over the A+B components only (magenta shaded region). In this case the line is fitted with a gaussian having a
FWHM=$178\pm 25$~km~s$^{-1}$, consistent with the one obtained from the global spectrum, and a peak intensity $F^{peak}_{[CII]}=47.6\pm 5.9$~mJy.
\begin{table*}
\caption{Properties of the BRI 0952-0115 components compared with optical observations by Leh\'ar et al. (2000): Right Ascension; Declination; Flux of each component not corrected for the lensing magnification factor. The uncertainties reported represent statistical errors. Absolute flux (systematic) errors are not relevant in this context; Flux normalized to component A (our observations); Flux normalized to component A (optical observations); Separation from component A (our observations); Separation from component A (optical observations)}
\label{tabcomp}     
\centering      
\begin{tabular}{c c c c c c c c} 
\hline\hline                
  &(J2000) RA&(J2000) Dec&$F_{[CII]}$& $(F/F_A)_{[CII]}$&$(F/F_A)_{opt}$&$(x-x_A)_{[CII]}$& $(x-x_A)_{opt}$\\
  &[sec]&[arcsec]&[Jy~km~sec$^{-1}$]&&&[arcsec]&[arcsec]\\
\hline                   
A& 09:55:00.125$\pm$0.002&-01:30:06.27$\pm$0.04&5.1$\pm$0.5& 1  &1 & 0&0\\
B& 09:55:00.088$\pm$0.003&-01:30:06.87$\pm$0.05&3.8$\pm$0.6& $0.75\pm 0.20$&$0.28\pm 0.01$&$0.82\pm0.06$&$0.993\pm0.003$\\
C& 09:55:00.021$\pm$0.006&-01:30:08.28$\pm$0.10&9.9$\pm$1.3& $2.00\pm 0.30$&--&$2.54\pm0.11$&--\\  
\hline
\end{tabular}\\
\end{table*}
\subsection{The lensed quasar host galaxy (component A+B)}\label{twocomp}

\begin{figure}
   \centering
   \includegraphics[width=9truecm]{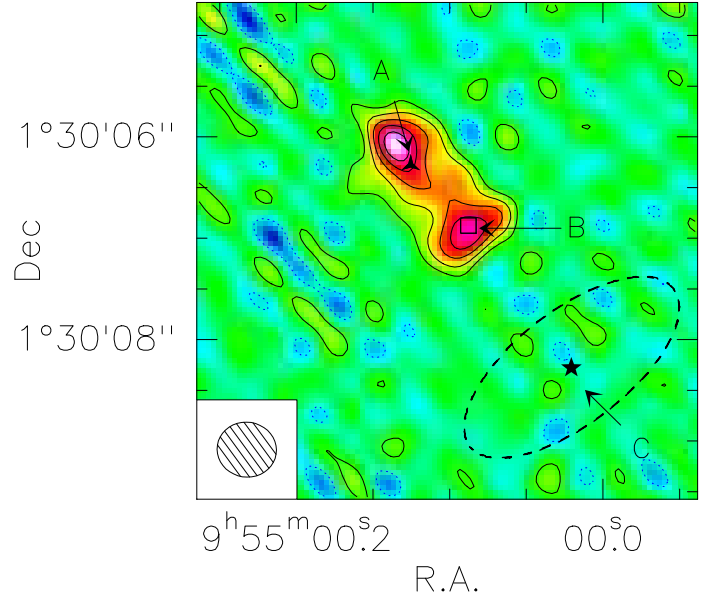}
      \caption{
Map of the [CII] emission of BRI 0952-0115 obtained with an inverted gaussian uv-taper to achieve an angular resolution of $0.5''\times 0.5''$. Contour levels are shown in steps of 1$\sigma=0.8$~Jy~km~s$^{-1}$~beam$^{-1}$. Negative contour levels at 1$\sigma$ are shown with dotted lines. The filled three-points star, empty square, filled five-points star, and the dashed ellipse have the same meaning as in the left panel of Fig. \ref{map}.}
         \label{mapzoom}
   \end{figure}

The structure A+B is most likely tracing star formation in the vicinity of the quasar nucleus. Moreover, the strongly elongated morphology of the structure is due to the blending of the double images lensed by the foreground galaxy. In fact, with the synthesized circular beam of 0.5$''$ (see Sec. \ref{secobs}) we can indeed resolve the A and B components, as shown in Fig. \ref{mapzoom}, while component C disappears since its extended emission requires short baselines contribution to be detected. For component A we estimate a radius of $\lesssim 0.2''$ (after deconvolving by the beam size), i.e. $\lesssim$~1.4~kpc.

We fit the component A+B in the uv-plane through a circular gaussian function with a $FWHM=0.2''$, finding the coordinates reported in Table \ref{tabcomp}. Although the beam is $\rm 1.08''\times 0.66''$, the signal-to-noise ratio (SNR) of this component is high enough that the analysis in the uv-plane allows us to determine the position of the two sources with an accuracy of $\lesssim \rm 0.1''$. In fact, the accuracy of a source position is given by $B_{synth}/(2\times SNR)$, where $B_{synth}$ is the synthesized beam size.

From the high resolution map of Fig. \ref{mapzoom} we note that the position of components A and B resulting from the fit in the uv-plane with all the baselines (filled three-points star and empty square, respectively) do not coincide with the peak of the [CII] emission. This may be due to the fact that the diffuse component of the [CII] emission has a centroid which is displaced with respect to the peak of the [CII] emission.

We obtain a separation of the lensed components A and B smaller than obtained through optical observations by Leh\'ar et al.\ (2000). In our observations, the component A appears to be separated by $\rm 0.82''\pm 0.06''$ from component B, instead of $\rm 0.993''\pm 0.003''$, as measured by Leh\'ar et al. (2000). Although small, the separation difference is significant relative to the uncertainties (Table \ref{tabcomp}).

We compute the flux associated to the A+B and C components by integrating over regions which follow the 2.5$\sigma$ contours on the map (see Fig. \ref{map}, left panel), covering $\sim 1.7~ \rm arcsec^2$ and $\sim 2.2~ \rm arcsec^2$, in the case of the A+B and C component, respectively. In particular, the relative flux of B with respect to A is the result of the fit to the A+B component. In Table \ref{tabcomp}, we report the flux associated with each component of the [CII] emission. One notes both that the [CII] emission is not coincident with the optical emission, and that the relative flux of B with respect to A resulting from our observations is significantly higher than in the case of the HST images. This is likely due to the fact that the [CII] emission (tracing star formation) is certainly more extended than the optical emission (coming from the accretion disk), hence the effect of differential magnification is reduced. Alternatively (or in addition) differential dust extinction from the lensing galaxy may also play a role. 
\begin{table*} \caption{Parameters of the fiducial lensing
model: redshift of the source ($z_S$); redshift of the lens ($z_L$);
mass of the lensing galaxy ($M_g$);
ellipticity ($\epsilon$); core radius ($\xi_c$); position angle ($PA$);
source position in the source plane relative to the lens galaxy
($x_S$,$y_S$); source size in the source plane ($R_S$); magnification factor of the source ($\mu_{\rm opt}$).}
\label{tablens}     
\centering      
\begin{tabular}{c c c c c c c c c c} 
\hline\hline                
 $z_S$&$z_L$&$M_g$&$\epsilon$&$\xi_c$&$PA$& $x_S$&$y_S$&$R_S$&$\mu_{\rm opt}$\\
& & $[M_{\odot}]$&& [kpc]& [degrees]&[arcsec]&[arcsec]&[arcsec]&\\         \hline   
4.43&0.632&$4\times 10^{10}$&0.065&0.025&284& 0.100$\pm$ 0.002& 0.095$\pm$ 0.002&0.07& 6.7$\pm$0.1\\
\hline
\end{tabular}\\
\end{table*}

\subsection{Lensing model}\label{lensec}
 \begin{figure*}
   \centering
   \includegraphics[width=16.truecm]{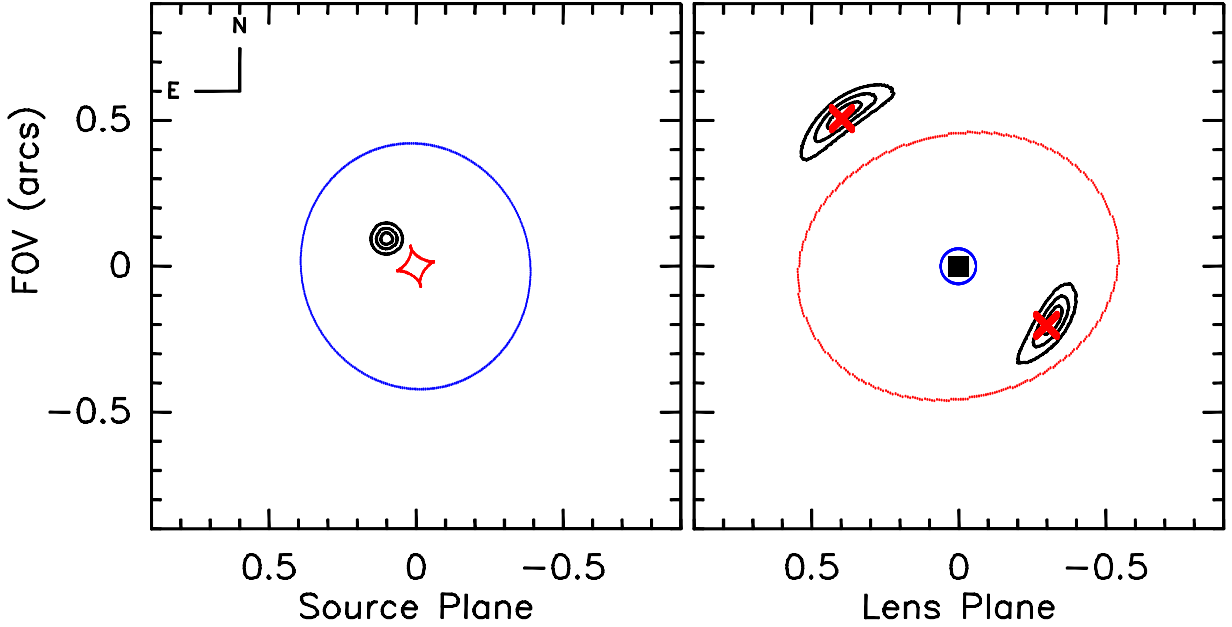}
      \caption{
Lensing model of BRI 0952-0115. Left panel: the lens galaxy (at z=0.632) is located at the center of the image. The black contours represent the BRI 0952-0115 position required to match optical observations. The mass distribution of the source is described in terms of a circular gaussian function with FWHM$\sim 0.07 ''$. The red curves around the position of the lens denote the tangential caustic, while the blue ellipse shows the radial caustic. Right panel: resulting image configuration and critical lines in the lens plane. The source is strongly lensed by the foreground galaxy and mapped onto a set of double images: the black contours represent the lensed images of BRI 0952-0115. Red crosses represent the relative position of A and B from the lensing galaxy location (filled red square), as measured by Leh\'ar et al. (2000) in the optical images. Contour levels are shown at 20\%, 50\%, and 80\% of the peak.} 
         \label{lensmodel}
   \end{figure*}
In this Section, we discuss the lensing model\footnote{We assume the
concordance $\Lambda$-cosmology with $\rm H_0=70.3~km~s^{-1}Mpc^{-1}$, $\rm
\Omega _{\Lambda}=0.73$ and $\rm \Omega _{m}=0.27$ \citep[][]{komatsu11}. With these cosmological parameters, at z=4.4336, an angular distance of 1$''$ corresponds to a proper distance of 6.85 kpc.} required to explain the structure of BRI 0952-0115 as revealed by optical observations and by the PdBI maps. For the mass distribution, we adopt the non-singular isothermal ellipsoid (NIE) model. We use the GLENS code by Krips \& Neri (2004). Eigenbrod et al. (2007) identified the lensing galaxy of BRI 0952-0115 as an early-type galaxy at $z=0.632\pm 0.002$.

The model has 7 free parameters: mass ($M_g$) of the lensing galaxy, its core radius ($\xi_c$, which represents the radius at which the mass density falls to half of its central value), ellipticity ($\epsilon$) of the projected mass distribution and its position angle (PA), source position relative to the lensing galaxy in the source plane ($x_S$,$y_S$); source size in the source plane ($R_S$). We assume that the mass distribution of the source is described in terms of a circular gaussian function with a $FWHM=0.07''$. First, we constrain the lensing model through the optical position of components A and B relative to the lensing galaxy in the lens plane (see Tab. 3 in Leh\'ar et al., 2000).

The results of the lensing model are shown in Fig. \ref{lensmodel}, where the panels show the source and the lens planes on the left and on the right, respectively. The tangential (red) and the radial (blue) caustic lines are displayed in the left panel. The corresponding critical lines are shown with the same color-code in the right panel. In the lens plane the two red crosses show the position of the optical images relative to the lensing galaxy (central red filled square), as measured by Leh\'ar et al. (2000), while the black contours represent the lensed images of BRI 0952-0115, as predicted by our model. The full set of model parameters is reported in Table \ref{tablens}.

Determining the errors on the lensing parameters is a long and demanding process, which goes beyond the scope of this paper, since we are not interested in the lens itself. In the context of this paper, the only parameter associated with the lensing model relevant for our discussion is the lens magnification. Hence, here we focus on the determination of the uncertainty of the lensing factor. We first compute a large grid over the lens shape parameters, and then, for the lensing models which provide a good agreement with the data, we calculate a grid over the source position to determine $\Delta \chi ^2<1$ ranges for the maximum and minimum values of the magnification factor. By varying the coordinates of the source position of $\sim 0.002''$ (see Table \ref{tablens}), we obtain $\rm \mu_{\rm opt}= 6.7\pm 0.1$ for the component A+B, while the C component is not gravitationally amplified. We note that the derived magnification factor does not strongly depend on the assumed size of the emitting region; in fact, by varying $R_S$ between one tenth of the fiducial value (i.e. 0.007$''$) and the radius upper limit obtained in Sec. \ref{twocomp} (i.e. 0.2$''$), $\mu_{\rm opt}$ changes only by 4\%. 

In Sec. \ref{twocomp}, we have noted that the distance between the [CII] components A and B appears to be smaller than that obtained by Leh\'ar et al. (2000), as shown in Table 2. In the case of mm observations, we do not know the relative position between the lens and components A and B. Therefore, we constrain the lensing model through the distance in the lens plane between components A and B, their flux ratio, and the position angle ($P.A.=40^{\rm o}\pm 2^{\rm o}$). We find that the lensing model which best describes the properties of the [CII] emitting regions differs from the previous one only in terms of the lensing galaxy mass, which in this case is 25\% smaller. We emphasize that in both lensing models the size of component A in the lens plane is smaller than 0.2$''$, in agreement with the radius upper limit found in Sec. \ref{twocomp}. By matching mm observations, we obtain a lower value of the magnification factor\footnote{A lower magnification factor can also be obtained by assuming an M/L +$\gamma$ model, as done by Maiolino et al. (2009).}, i.e. $\mu_{\rm mm}=5.8\pm 0.7$, which is however still consistent within 1.3$\sigma$ with $\mu_{\rm opt}$.\\ 
Further efforts are required to investigate the properties of the A and B components in more details. In a forthcoming paper, we will present higher SNR and higher angular resolution observations of the CO emission in BRI 0952-0115. For the time being, we consider as fiducial the amplification factor constrained through optical observations, which is characterized by a smaller uncertainty.
\subsection{The companion galaxy (component C)}

The C component is more extended than the component A+B. We cannot completely rule out the
possibility
that the C component may result from a residual sidelobe due to a sub-critical coverage of the
uv-plane; however, we regard this as very
unlikely, since region C is detected at a confidence level of $\sim 7\sigma$.

We fit the C component with an elliptical gaussian function in the uv-plane, finding a size of the emitting region of $2.5''\times 1.1''$ ($\sim 17\times 7$~kpc)
centered at the coordinates reported in Table
\ref{tabcomp}. The filled five-points star and dashed ellipse in Fig. \ref{map} (left panel) show the results of the fit,
while, as mentioned, component C is not detected in the high resolution map of Fig.~4, because this map
misses the short baselines sensitive to emission more extended than 0.5$''$.

We note that component
C is not detected in HST images, probably because of its low surface brightness or high extinction,
which makes it undetectable in the optical.
The published CO(5--4) map
\citep{gui99} cannot help to constrain the nature of the C component, since
the angular resolution of these observations is too low ($6''\times 5''$) to disentangle
the components A+B and C. VLA observations at higher angular resolution ($\sim$ 2")  suggest the presence of CO(2--1) emission extended over about 2.5" South-West with respect to the optical quasar positions (Carilli et al. in preparation). However, the SNR of these maps is not sufficiently high to obtain any definitive conclusion.

\begin{figure*}
   \centering
\includegraphics[width=18cm]{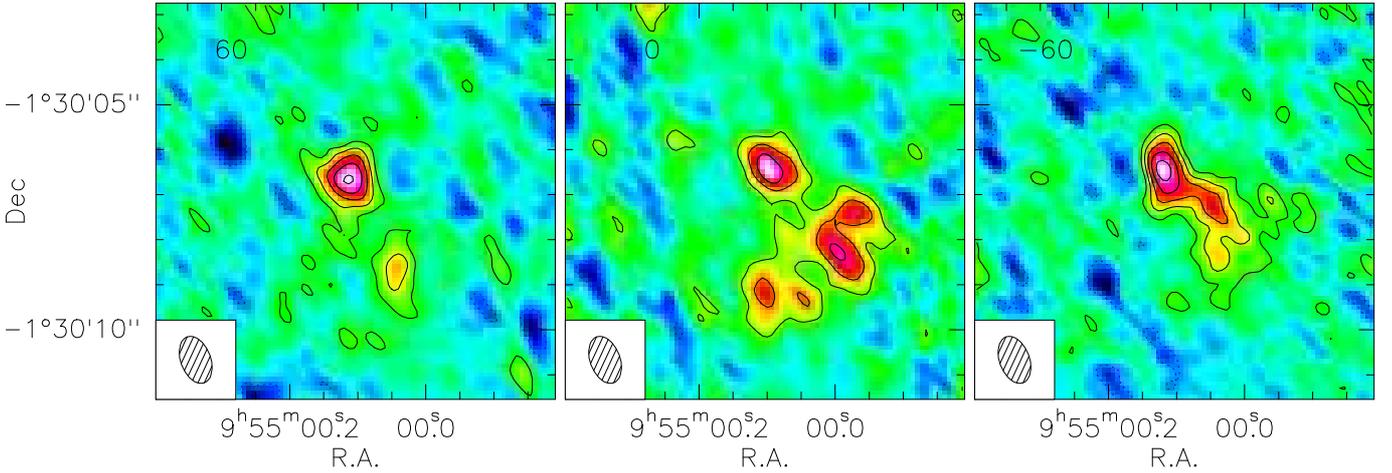}
      \caption{
Maps of the [CII] emission obtained with the PdBI in 60 km~s$^{-1}$ wide channels. The central velocity of each map (in km~s$^{-1}$) is indicated in the upper left corner in units of km~s$^{-1}$. Contour levels are shown in steps of 1.5$\sigma$, where 1$\sigma=0.2$~Jy~km~s$^{-1}$~beam$^{-1}$. Negative contour levels of 1.5$\sigma$ are represented through dotted lines. Source C reveals a complex velocity
pattern, suggestive of rotation disturbed by the interaction with the quasar host galaxy (A+B).}
         \label{gradvel}
   \end{figure*}

The [CII] line channel maps (Fig. \ref{gradvel}) show a complex and irregular velocity field, not easy to interpret
in terms of simple galaxy rotation. Most likely, component C is a companion disk galaxy
whose velocity field is distorted by the interaction with the quasar host galaxy.
\subsection{SFR and $\rm \Sigma _{SFR}$}
We estimate\footnote{We define the far-IR luminosity as $\rm L_{FIR} = L(42-122)~\mu m$ (Stacey et al. 2010).} $\rm L_{FIR}$ by using the galaxy template by Polletta et al. (2007) of a local ULIRG/QSO (Mrk231) normalized to the continuum observed by us at $\lambda_{rest}\sim 158~\mu \rm m$. Note that since we probe the FIR continuum close to the peak of the SED, the inferred $\rm L_{FIR}$ is not strongly dependent on the adopted template (i.e. on the average dust temperature).
For the A+B component, the measured continuum ($F_{cont}=11.4\pm 1.4$~mJy) and the magnification factor determined through our fiducial lensing model ($\rm \mu_{\rm opt}= 6.7\pm 0.1$) provide a de-lensed luminosity $\rm L_{FIR}^{A+B}=(1.5\pm 0.3)\times 10^{12} \rm L_{\odot}$. If
$\rm L_{FIR}$ is associated with star formation, as in most quasars, even at high redshift (Lutz et al., 2008), then this
luminosity corresponds to a star formation rate $\rm SFR=270\pm 40~M_{\odot}~yr^{-1}$ \citep{kenn98}.

As far as component C is concerned, since the continuum is only barely detected, we conservatively use the 3$\sigma$ upper limit on the FIR continuum (i.e. 4.2 mJy). We obtain $\rm L_{FIR}^C < 3.8\times 10^{12} \rm L_{\odot}$, which corresponds to $\rm SFR < 660~M_{\odot}~yr^{-1}$.

Taking into account the upper limit on the size of component A mentioned in Sec. \ref{twocomp}, i.e. $\lesssim$~1.4~kpc, and assuming that component A contributes half of the total FIR flux, we infer a SFR surface density of
$\rm \Sigma _{SFR}^{A+B}\gtrsim 
150~M_{\odot}~yr^{-1}~kpc^{-2}$. Such a high rate of star formation per unit area is typical of the active star forming region in local ULIRGs (Siebenmorgen et al., 2008; Elbaz et al., 2011).

The compactness of the star formation in the host of the quasar nucleus is confirmed by the
$\rm L_{[CII]}/\rm L_{FIR}$ ratio.
This ratio is considered to be a tracer of the radiation field density and therefore a proxy for the compactness
of star formation (Stacey et al., 2010).
For the A+B component, we obtain $\rm (L_{[CII]}/\rm L_{FIR})^{A+B}=(5.3\pm 1.1)\times 10^{-4}$, typical of local compact
ULIRGs (Luhman et al., 1998; Luhman et al., 2003). Unfortunately, a resolved CO map is still not available to isolate the CO emission of component A+B from component C
and, therefore, prevents us from locating the individual components on the $\rm L_{[CII]}/\rm L_{FIR}$ versus
$\rm L_{CO}/\rm L_{FIR}$ diagram, which would allow us to better model the physical conditions of the star forming
regions in these components.

We emphasize that the calculations of the SFRD and [CII]/FIR ratio do not depend on the magnification factor. By assuming two extreme values of the magnification factor ($\mu=3.7$ and $\mu=7$, i.e. the 3$\sigma$ lower limit of $\mu_{\rm mm}$ and the the 3$\sigma$ upper limit of $\mu_{\rm opt}$), the mean de-lensed FIR luminosity gives a star formation rate $150\leq \rm SFR\leq 280~\rm M_{\odot}~yr^{-1}$.

In the companion galaxy C, we can only infer an upper limit on the surface density of star formation.
Based on the upper limit on the SFR and the size measured in the previous section, we infer
$\rm \Sigma _{SFR}^{C}\le 5~M_{\odot}~yr^{-1}~kpc^{-2}$, which is consistent with moderate intensity local
starburst galaxies (e.g. M82, NGC253, Arp299). The spatial extent of the star formation is also supported by
the $\rm L_{[CII]}/\rm L_{FIR}$ ratio. More specifically, for the companion galaxy we obtain $\rm (L_{[CII]}/\rm L_{FIR})^C \gtrsim 2\times 10^{-3}$, again typical of extended starburst galaxies.

This indicates that the interaction between two galaxies at high-z can boost both compact star formation and
black hole accretion in one of the galaxies, while enhanching star formation over the whole
disk of the other.

\section{Conclusions}

We have presented one of the first resolved maps of [CII] 158$\mu$m emission at high redshift. By exploiting the new IRAM PdBI receivers at high frequency we have
observed BRI 0952-0115, a lensed quasar at z=4.4, that is characterized by strong [CII]
emission. The PdBI map reveals a surprisingly complex structure.
The [CII] emission can be divided in two main components: a compact structure
(A+B), associated with the two lensed optical images of the quasar,
and a second more extended component located 2$''$ South-West of the quasar.

The [CII] emission associated with the quasar (component A+B) is clearly
resolved in two lensed images as the optical quasar images. The
[CII] emission associated with the quasar nucleus is distributed over a region of 1.4 kpc or less.
The continuum is also detected in this component, with a de-lensed luminosity of
$\rm L_{FIR}^{A+B}=1.5\times 10^{12}~\rm L_{\odot}$, implying $\rm SFR\sim 270~ M_{\odot}~yr^{-1}$ .
The inferred SFR surface density is $\rm \Sigma _{SFR}\gtrsim 150~ M_{\odot}~yr^{-1}~kpc^{-2}$. Such a high
rate of star formation per unit area is similar to that observed in local ULIRGs (Siebenmorgen et al., 2008; Elbaz et al., 2011). The compactness of the star formation
in the host of the quasar BRI 0952-0115 is also supported by the ratio
$\rm (L_{[CII]}/\rm L_{FIR})^{A+B}\sim 10^{-4}$, characteristic of high UV radiation fields
typically associated with compact ULIRGs.

We also detect another [CII] emitting object (C), located about 2$''$ South-West of the A+B component. The [CII] emission of this component is clearly resolved over about 12~kpc. This is likely a companion galaxy, located at about 10~kpc from the quasar, in the process of merging with the quasar host. The velocity pattern of this component is irregular, suggesting a rotation field distorted by the tidal interaction with the quasar host galaxy. The continuum is only barely detected and indicative of a SFR density ($\rm \lesssim 5~M_{\odot}~yr^{-1}~\rm kpc^{-2}$) lower than in the quasar host galaxy and more typical of extended starbursting disks. The extended and diffuse nature of star formation in this companion galaxy is further supported by the high $\rm L_{[CII]}/\rm L_{FIR}$ ratio ($\gtrsim 2\times 10^{-3}$).  

Galaxy mergers may have quite different effects on the interacting galaxies. In the case of BRI 0952-0115 the galaxy interaction has both boosted compact star formation in the central region and black hole accretion in one of the two galaxies, while in the other galaxy the interaction has increased star formation all over the galaxy disk.

These results highlights the power of [CII] mapping to investigate the nature of quasar host galaxies and, more generally, star formation in high redshift galaxies.
\begin{acknowledgements}
This work was done while SG was visiting the Institut de RadioAstronomie Millimetrique and the Plateau de Bure Interferometer in Grenoble, whose hospitality is warmly acknowledged. SG is grateful to R. Cesaroni, F. Fontani, Y. Libert, and A. Mignano for useful discussions during the reduction procedure of the PdBI data. We thank the anonymous referee for her/his useful comments and suggestions. 

\end{acknowledgements}

\end{document}